# Erwin Schroedinger, Francis Crick and epigenetic stability

Vasily V. Ogryzko

Université Paris-Sud 11
CNRS, Interactions moléculaires et cancer UMR 8126
Institut de Cancérologie Gustave-Roussy
Villejuif, France

Email address: vogryzko@gmail.com

# Abstract

Schroedinger's book 'What is Life?' is widely credited for having played a crucial role in development of molecular and cellular biology. My essay revisits the issues raised by this book from the modern perspective of epigenetics and systems biology. I contrast two classes of potential mechanisms of epigenetic stability: 'epigenetic templating' and 'systems biology' approaches, and consider them from the point of view expressed by Schroedinger. I also discuss how quantum entanglement, a nonclassical feature of quantum mechanics, can help to address the 'problem of small numbers' that led Schroedinger to promote the idea of a molecular code-script for explaining the stability of biological order.

# Reviewers

'Molecular biology has been successful largely because it has concentrated on the type of problem […] that can be attacked by isolating a small part of a biological system'
F. Crick, 'Molecular Biology in the Year 2000'

"Living matter, while not eluding the 'laws of physics' as established up to date, is likely to involve 'other laws of physics'… It is, in my opinion, nothing else than the principle of quantum theory over again"
E. Schroedinger, 'What is Life?'

In one of the most influential books on science in 20th century, 'What is Life?', Erwin Schroedinger, a founder of Quantum Mechanics, asked what physical principles govern the stability of biological systems. He suggested that the physics of the covalent bond holds the key to the secret of heredity and popularized the idea of DNA representing a molecular code-script for the genetic makeup of an organism. The book inspired the collaboration between Crick and Watson and led to the identification of genetic information as the sequence of bases in DNA, replicating via complementary base-base recognition.

The recent surge of interest in 'all things epigenetic' shows that the issue of stability, let alone heredity, of biological systems is far more complex than one could envision half a century ago. While recognizing the role of DNA sequence as the dominant contributor to the persistence of biological order, the emerging view offers a richer spectrum of additional factors that contribute to biological organization in a manner somewhat independent from DNA. A large subset of these factors, summed up under the name of epigenetic information, is responsible for the maintenance of differentiated cell types in development, plays an important role in cancer and has been making headlines lately as the culprit responsible for the setbacks in reproductive cloning. Could the insights of Schroedinger's little book into the physics of biological order be useful in sorting out the nature and potential mechanisms of processing, storage and transmission of epigenetic information?

For Schroedinger, the most obvious manifestation of biological stability was heredity, hence his interest in the physics of inheritance. Likewise, current epigenetic studies also emphasize heritable aspects of epigenetic phenomena. This focus, however, unnecessarily limits the scope of epigenetics. Terminally differentiated cells live for decades, maintaining their distinct phenotypic traits in spite of environmental stresses, thermal noise and DNA damage/repair. These cells do not proliferate, thus the term 'heritable' does not apply to their traits. The term *'epigenetic stability'* refers to a broader phenomenon that encompasses maintenance of phenotypic traits in both nonreplicating and replicating cells independent from DNA sequence. In this essay, I contrast two classes of potential mechanisms of epigenetic stability and discuss them from the point of view enunciated by Schroedinger. Whereas the first class, termed here 'epigenetic templating', does not depart far from the molecular biological perspective on life, an analysis of the 'systems biology' mechanisms might lead to far-reaching changes in our understanding of the physical basis of information processing in living cells. Interestingly, Schroedinger's ideas remain relevant in this new development.

*Francis Crick and 'epigenetic templating'*

Although the existence of epigenetic mechanisms conflicts with the 'genocentric view of life', it does not directly challenge the Central Dogma of molecular biology, which clarifies how genetic information is deciphered to give rise to biological organization. Francis Crick formulated the Dogma carefully as forbidding 'the detailed *residue-by-residue* transfer of sequential information' from proteins to nucleic acids or other proteins [1]. This definition leaves the door open for other types of information that could be required to specify the state of the organism (e.g. stored in macromolecular conformations, interactions and post-translational modifications) and might propagate independently from the DNA sequence.

Moreover, it was Crick who, later in his career, became preoccupied with the molecular mechanisms of the essentially 'epigenetic' phenomenon of neurobiological memory. Crick did not believe that the strength of an individual synapse, often located far from the nucleus, could be encoded in a nucleic acid, and instead favored a protein modification as the effector. But how then can the synaptic strength (the elemental basis of memory) persist for decades, in spite of the molecular turnover that tends to erase, in a matter of days or weeks, the molecular records stored in protein structures?

Crick postulated that i) the strength of a synapse is determined by phosphorylation of a protein molecule, ii) this protein can form dimers (or oligomers) and iii) the kinase responsible for the modification will only modify monomer in a dimer that already has the other monomer phosphorylated. This model of kinase action (inspired by the semi-conservative mechanism for the maintenance methylation of DNA [2]) would allow the cell to restore the overall phosphorylated state of the dimer/oligomer whenever molecular turnover replaces either monomer by a new one. The result is an extended half-life of the structure, i.e., memory, as 'the molecules in the synapse […] can be replaced with new material, one at the time, without altering the overall state of the structure [3]'.

A recent proposal of the involvement of a self-perpetuating prion-like mechanism in the persistence of long-term memory [4] echoes Crick's thought. More importantly, and regardless of whether such a mechanism plays a role in the long-term synaptic potentiation, it has wider applicability, not restricted to the memory mechanisms and protein phosphorylation. A case in point is chromatin, considered now as a principal carrier of epigenetic information. In addition to DNA methylation, the maintenance of genome-wide patterns of post-translational histone modifications in differentiated cells has been proposed to operate via similar self-perpetuating mechanisms. For example, histone acetyl-transferases (HATs) often contain bromodomains that recognize acetylated lysines. This design should facilitate maintenance of acetylated states of chromatin by recruiting HATs to acetylated histones. On the other hand, histone methyltransferases are often linked with chromodomains, the recognition modules for (tri)methylated lysines, suggesting a similar feedback loop for this histone modification [5-7]. In absence of a better word, we used the term *'epigenetic templating'* [8] to refer to a mechanism of perpetuation of epigenetic information that is based on the preferential activity of enzymes that deposit a particular epigenetic mark on macromolecular complexes already containing the same mark (Fig. 1).

The recent proposal of a semi-conservative mechanism of nucleosome duplication [9] might mean that epigenetic templating also works for chromatin marks of a different kind, e.g., variant histones. All variants of histone H3 form so called homotypic nucleosomes, having both H3 molecules of the same type in the histone octamer. However, new H3 histones are deposited on DNA only in the form of H3/H4 dimers [9]; thus the deposition machinery must match the types of new and old H3 histones if it is to keep the nucleosome homotypic. Such a mechanism could contribute to an understanding of the role of CenpA, the centromere-specific version of histone H3, in the epigenetic maintenance of centromere structure independent of DNA sequence [10].

Intriguingly, the design of some protein kinases is suggestive of epigenetic templating also operating in the cytoplasm. Tyrosine kinases often contain SH2 domains, which recognize phosphotyrosine [11], whereas some serine/threonine kinases contain FHA domains that recognize phosphoserine/phosphothreonine [12]. These structural features are usually understood in terms of the signaling cascade paradigm, with a phosphoprotein A modulating a protein kinase targeting a downstream protein B. However, if some of the kinase targets form dimers (or oligomers), they can perpetuate their phosphorylation status very much in line with the Crick's original suggestion. With many kinases residing in the cytoplasm, the associated epigenetic information would have to be classified as *trans-acting*, i.e., not marking any particular genomic locus. This choice of terminology would help to distinguish it from the chromatin-associated, *cis-acting* epigenetic information, which comprises DNA methylation, post-translational histone modifications, alternative histones etc.

A shift from the signal transduction paradigm to the epigenetic one might be useful for the study of many other protein modifications and even of interactions between macromolecules. Suppose that we are looking for a protein domain recognizing an ADP-ribosylated lysine. The possibility that a certain ADP-ribosylase could be an enforcer of epigenetic templating will instruct us that this very enzyme should be the first place to look for the ADP-ribosylated lysine recognition domains - an idea that is rather unexpected from the signal cascade point of view. Likewise, if epigenetic information can be stored in macromolecular interactions (exemplified by alternative histones binding to and marking a particular segment of DNA), the corresponding chaperones facilitating these interactions might be designed to enforce epigenetic templating, as considered in [8]. Future studies should reveal to what extent this model, promising to complement the Central Dogma of molecular biology, could explain the stability of epigenetic phenomena.

*'Systems biology' approaches to epigenetic stability and the problem of fluctuations*

That Francis Crick, the father of Central Dogma, entertained ideas extending beyond its boundaries, is not surprising and serves to illustrate that the better you know a concept, the more you see its limitations. Yet, although at odds with the genocentric view of life, the epigenetic templating idea remains true to the general spirit of molecular biology, which, following Schroedinger, looks for the keys to life in molecular structure. The model shares with the Watson-Crick mechanism of DNA replication one essential feature in common - the localized 'bit by bit' character of information storage and transfer. Information is stored in a particular macromolecular state (e.g., protein modification) and its transfer relies on inter-molecular contacts and physical linkage between different (modification-deposition and modification-recognition) activities. Clearly, such a model has the advantage of being testable by traditional biochemical experiments in cell-free systems. However, is it the only game in town?

Recent advances in 'omics' technologies have opened a way to analyze life from a different perspective, that of systems biology. The rediscovery that life is, in fact, a complex system phenomenon leads to radically new approaches to explain the stability of biological order based on the global dynamic properties of living systems. For example, the ability of gene regulatory networks to have multiple alternative attractors was long proposed to account for the stability of different patterns of gene expression during cell differentiation [13]. A more general, and related, view compares the organization of intracellular structure and dynamics to the order spontaneously generated in thermodynamic systems far from equilibrium [14, 15]. Unlike in the case of 'epigenetic templating', the epigenetic factors that specify cellular organization in these 'systems biological' approaches cannot be traced to a particular molecular structure, or to some other 'small part of a biological system'. Rather, this kind of epigenetic information has a global status, representing one choice out of a spectrum of possible stationary regimes in the dynamics of the whole system.

Given the growing appreciation of systems approaches to many aspects of life, how plausible are the

'systems biology' approaches to epigenetic stability? The harsh reality test facing them is the intrinsic noisiness of the intracellular dynamics due to the small numbers of participating molecules. The fluctuations in the numbers of molecules of transcriptional regulators render questionable the role of gene network dynamics in stability of gene expression patterns [16, 17](although the stochastic nature of genetic circuits is likely involved in making *choices* between these patterns during differentiation). The 'small numbers curse'[1] looms equally when the concepts from non-equilibrium statistical physics and classical chemical kinetics, which work reasonably well on a macroscopic scale, are applied to the organization of such nano-objects as bacterial cells or eukaryotic cells' compartments, having only a few copies of many vital molecular components [18].

Exacerbating the 'small numbers' problem are two factors: the great variety of different molecular species present in even the simplest living cell and the need to take into account the local context of each molecule. Both lead to a very large number of variables necessary to describe the state of a cell. To see how the problems of fluctuations and high dimensionality of intracellular dynamics combine, consider the following toy example. A single *E. coli* cell contains 2000 molecules of RNA-polymerase enzyme [19]. As a measure of the stability of a particular variable $r$, let us take the cost of its maintenance ($Mr$), i.e., energy needed to be dissipated in order to keep the value of $r$ in a range acceptable for the proper functioning of the cell. Intuitively, the more $r$ fluctuates, the costlier it will be to maintain its value within the acceptable range. In the context of the 'small numbers problem', $Mr$ is proportional to $\sqrt{Nr}$, where $Nr$ is the number of molecules of the $r$ species in a single cell, and $\sqrt{Nr}$ is the average fluctuation size. If we first suppose that the value of 2000 molecules is all we need to know about the state of RNA-polymerase *in vivo*, then the average size of the fluctuation is $\sqrt{2000} = 45$, i.e., only 2% of the total number of RNA-polymerase molecules. We might then conclude that the cost $Mr$ of maintaining the proper state of this enzyme *in vivo* is negligible, as this variable does not appear to be affected much by the fluctuations. But consider now the evidence of functional differentiation of RNA polymerase in *E. coli*, attributed to its post-translational modifications, existence of several species of σ subunit and 100–150 transcription factors affecting the promoter selectivity and mode of action of this enzyme [19]. Overall, this evidence suggests that to know accurately the functional state of RNA-polymerase in a single cell, every molecule of RNA-polymerase might have to be counted as a separate variable. With this increased resolution of our description the troublesome effect of fluctuations also grows, and quite dramatically, as the cost of keeping the proper state of RNA-polymerase *in vivo* (now many variables $r_i$ instead of a single one) becomes proportional to $Nr$ instead of $\sqrt{Nr}$ (since if the number of labels $i = N$, the $\Sigma\sqrt{Nr_i} = Nr$ due to $\sqrt{1}=1$).

If molecular biology tells us anything, it is that the more we learn of the intricacies of intracellular regulation, the more functional differences between different molecules of the same species we find, due to their many modifications and their placement in particular intracellular contexts. If, in the spirit of the 'omics' approach, we consider all this information to be essential for the complete description of intracellular dynamics, the number of variables will expand dramatically, together with the cost of keeping fluctuations under control. This reflects the fact that increasing the resolution in our description of the cell state is the reverse of 'coarse graining', a procedure used in statistical mechanics to transit from a detailed 'microscopic' description of a physical system to its description in terms of 'macroscopic' degrees of freedom by averaging out all nonessential variables. Thus, the more information about the cell we consider as essential, the less we can rely on the law of large numbers to account for the stability of intracellular dynamics.

Could the lessons of Schroedinger's book help to get us out of trouble? In fact, the very same 'small numbers' problem was exactly his reason for believing that 'new physics' should be involved in explaining the stability of biological order, and for promoting molecular structure as the basis for genetic inheritance. His logic was as follows. Whereas the stability of most of the macroscopic phenomena is based on the averaging of properties of large numbers of participating elements behaving chaotically at the micro-level, the molecules are held together by the laws of quantum mechanics, and their structure and composition are impervious to statistical fluctuations. Quantum principles, therefore, would have to have direct relevance to the stability of biological order, where, in manifest contrast with thermodynamic systems, 'incredibly small groups of atoms, much too small to display exact statistical laws, do play a dominating role in the very orderly and lawful events within a living organism' [20].

This analysis was impressively confirmed by the ensuing progress of molecular genetics. It established beyond any doubt the principal role of the covalent sugar-phosphate DNA backbone in the ability of biological systems to carry arbitrary amounts of genetic information. As discussed above, the idea of molecular structure as the basis for information storage is equally consistent with the 'epigenetic templating' model. But so far as epigenetic information is concerned, could quantum principles also contribute to the 'systems biology' approaches to epigenetic stability, similarly helping them to deal with the 'small numbers problem'?

*Schroedinger and quantum entanglement*

At first glance, appealing to quantum mechanics to solve the fluctuation problem only seems to complicate the situation. In quantum theory, the dimensionality of a composite system does not grow linearly with the number of its parts (as it does in the classical case), but exponentially [21]. Thus, as the number of variables used to describe intracellular dynamics multiplies, the *Mr* needed to keep all the fluctuations under control has to grow exponentially. Would not, then, appealing to quantum mechanical principles in explaining intracellular dynamics amount to shooting ourselves in the foot? Would it be wiser of us to restrict the biological applications of quantum theory to the time-honed studies of molecular structure and intermolecular interactions only? This does not seem to be the best strategy. Quantum theory is a general theory of the stability of matter, indispensable for understanding the behavior of objects both small and large. Being a particular case of condensed matter physics, it is very unlikely that the physics of life can be properly understood without quantum theory. As we perfect our technology to the highest resolution, the description of intracellular dynamics starting from first (i.e., quantum mechanical) principles will become increasingly important. Also, the ongoing progress in 'omics-' and 'nano-' technologies will eventually lead us to an understanding of the limits to how much can be observed concerning an individual biological object (e.g., single cell). A natural language to take these limits into account could be the formalism of quantum theory. However eccentric it may seem, such a line of inquiry promises to bring us some surprises in our understanding of life [22, 23].

*Quantum entanglement* (Fig. 2), perhaps the most non-classical feature of modern physics, might be most relevant in taking the leap to the quantum mechanical description of biological systems. The last several decades of experimental and theoretical work demonstrate that it is a universal property of physical systems composed of interacting parts [24]. It manifests itself experimentally in correlations between spatially separated events that are not causally related [25], thus providing physics with a new kind of order, unexpected from the classical worldview. In addition to the proposals to use it as a radically new resource for information processing[2] [26], recent work provides increasing theoretical and empirical evidence that entanglement can be surprisingly robust and can exist at high temperatures in material systems [27, 28]. My expectation is that the physical description of intracellular processes from first principles will require taking entanglement into account. The systems biology approaches based on approximations that get rid of entanglement at the outset will have serious difficulties explaining the stability of intracellular dynamics.

How may entanglement substitute 'large numbers' in stabilizing biological order? Similar to the coarse graining procedure, it can reduce the number of degrees of freedom that have to be protected from fluctuations, but in a more subtle way than by simple averaging. Consider the simplest example of an entangled system – so called Bell pair of elementary particles [26]:

$$|\Psi\rangle = (|\uparrow\rangle_1|\downarrow\rangle_2 + |\downarrow\rangle_1|\uparrow\rangle_2)/\sqrt{2}, \qquad (1)$$

where the arrows designate the spin orientation of particles 1 and 2,
and compare it with a non-entangled pair (product state):

$$|\Psi\rangle = |\rightarrow\rangle_1|\rightarrow\rangle_2 = ((|\uparrow\rangle_1 + |\downarrow\rangle_1)/\sqrt{2})((|\uparrow\rangle_2 + |\downarrow\rangle_2)/\sqrt{2}) =$$
$$= (|\uparrow\rangle_1|\downarrow\rangle_2 + |\uparrow\rangle_1|\uparrow\rangle_2 + |\downarrow\rangle_1|\uparrow\rangle_2 + |\downarrow\rangle_1|\downarrow\rangle_2)/2 \qquad (2)$$

Whereas, in the 'spin up'/'spin down' representation, all combinations of the states of the two particles are allowed for the product state (2), in the case of the Bell pair (1) only two are. It is manifested in correlations in the results of measurements of the spins of these particles. In an extreme case, when measured in the 'spin up'/'spin down' basis, the Bell pair behaves as if it has only one degree of freedom (instead of two, as expected from its composition); in this representation, *entanglement effectively reduced the number of dimensions* of this composite system. Accordingly, with an increase in the number of components, the number of independent variables essential to describe an entangled system might not grow as fast as in the case of a separable system.

Now, going back to systems biology, we can ask whether entanglement has to be taken into account when choosing proper variables for the description of intracellular processes. The description of the cell state from first principles has to start with specifying the position of every nucleus and electron in it[3]. The question then becomes how to simplify this overwhelmingly complicated picture. We saw that there are limits to how far we can go with the coarse graining procedure without losing essential information about the cell's functioning. On the other hand, interactions between different parts of the cell will inevitably entangle them, and we also saw that discarding information about the entanglement will artificially increase the number of independent degrees of freedom, with a concomitant growth in the cost of maintenance[4]. These arguments suggest that the proper description of the cell's functioning will require taking some of this entanglement into account, in order to keep the expected total cost of maintenance under control. The resultant acknowledgment of correlations between molecular events inside the cell can help to understand how intracellular dynamics remains robust in spite of the fluctuations in the numbers of its many participating components.

Finding support for the proposed role of entanglement meets with both experimental and theoretical challenges. First, we will need to learn how to detect correlations in the fluctuations of many observable characteristics of living cells, and to be able to do it on a *single* cell level. This task is still difficult with current technology and will require the marriage of 'nano-' and '-omic' approaches, still in their infancy. Second, entanglement is commonly perceived to be very fragile and to be quickly destroyed due to the interaction of the physical system with its environment. This phenomenon, called environmentally induced decoherence (EID), has hindered the otherwise remarkable development of protocols that use entanglement for more efficient computation and secure cryptography [24]. There are grounds for optimism, however. The perceived fragility of *any* entanglement is a misconception resulting from the known challenges in controlling a particular class of entangled states for the purposes of quantum information processing. In fact, entanglement is a ubiquitous phenomenon, always present whenever there is an interaction between different physical systems; even the EID itself is a consequence of the entanglement of a system with its environment. The language that explicitly includes entanglement is bringing more clarity, convenience and additional insights into the study of known physical phenomena (see, for example [29, 30]). It is acquiring increasing importance in explaining the properties of condensed matter [31, 32], leading to speculations on its crucial role in life [33].

Indeed, if it has taken Humankind only few decades to approach the use of entanglement in quantum information technology, one can wonder why Life, in billions of years of evolution, could not also learn to take advantage, finding in entanglement an alternative resource for stabilizing biological order. Could it have learned long ago to run intracellular dynamics in so-called 'decoherence free subspaces' [34, 35], recently discovered by quantum information theorists in their efforts to protect quantum information processing from EID? Quite strikingly, a selection process might have been involved here, but of a kind that is somewhat different from and could precede the canonical Darwinian mechanism. Mathematically,

'decoherence free subspaces' are related to the concept of 'preferred states' proposed to explain transition from the quantum world to the classical [36]. In this approach, EID does not destroy all entangled states - some are stabilized via a process called environmentally-induced superselection. Accordingly, the entangled states in biological systems could be protected from EID exactly because they are the 'preferred states' *surviving* interaction with the environment. The implications of this idea could be interesting for both information technology and the understanding of life. It suggests a perspective on the adaptation of life to its environment that differs from the canonical Darwinian view in that adaptation via 'survival of the fittest' can happen on the level of an individual object, i.e., does not involve replication [22, 23]. On the other hand, it indicates a new potential source of entangled states prepared for us by biological evolution, which could be used in quantum information protocols.

To summarize, entanglement can provide biology with a conceptual ingredient that had been missing from the molecular explanations of life dominating the field for the last 50 years. A philosopher could see in it the physical counterpart of an old dictum – 'the whole is more than a sum of its parts', reflecting the aspect of life that cannot be reduced to molecular structure and interactions. A modern information theorist would see entanglement as an independent resource for information processing in living cells, additional to the molecular 'nuts-and-bolts' mechanisms (including epigenetic templating), which would be tempting to relate with the LOCC operations in this context. Incidentally, it was also Schroedinger who first recognized entanglement as '*the* characteristic trait of quantum mechanics, the one that enforces its entire departure from classical lines of thought' [37]. Quite ironically, then, the legacy of the Schroedinger's quest might outlast the molecular revolution in life sciences that it helped to initiate, and take us to a new understanding of biological organization - from the systems perspective, assimilating along the way the spectacular advances of modern physics and information science.

## Reviewers' comments

Eugene Koonin, NCBI, NLM, NIH, Bethesda, MD, USA

    This is a very appealing, actually, exciting essay. Ogryzko turns to Schroedinger's classic book and also to Crick's ideas on the mechanisms of synaptic memory in an attempt to outline two major mechanism through which cells could maintain epigenetic stability.
    The first mechanism is the very straightforward, even if elegant, "epigenetic templating", a sort of autocatalytic self-perpetuation of macromolecular modification. This is, in my opinion, a completely sensible, readily testable hypothesis. In addition to direct experimental tests, I believe, it is possible to

shed more light on the importance of epigenetic templating by a genome-wide computational analysis of protein domain architectures.

The second facet of epigenetic stability discussed here is completely different in character and decidedly non-orthodox. In essence, Ogryzko proposes that the "small numbers curse", which is at the center of Schroedinger' treatise, i.e., the way a cell copes with disruptive effects of molecular fluctuations, could involve quantum entanglement, the famed non-local aspect of quantum physics. The fluctuations problem is real and troubling, and the proposed solution seems to be ingenious and, potentially, powerful (caveat emptor: my own understanding of entanglement is at the level of News & Views articles). The idea of stabilization of particular entangled states via environmentally induced superselection is particularly attractive. In principle, this could be "it", the Holy Grail of Biological Physics, a non-trivial (not limited to properties of molecules) role of quantum effects in biological systems. Of course, it must be clearly realized that the proposal outlined in this paper is not a theory and not even a full-fledged hypothesis: it is "just" an idea but, I think, one that biologists cannot afford to dismiss.

The historical context of the paper, with all the multifaceted ramifications of Crick's and Schroedinger's classics, might not be exactly essential for the presentations of Ogryzko's ideas, but this surely makes for enticing reading. The only aspect of the paper that I do not like is the use of the "systems biology" jargon in the second part. It requires thinking but there surely must be a better way to describe the effects of fluctuations on cell functioning.

*Author's response – I thank Dr. Koonin for his comments. I appreciate the idea of using genome-wide computational analysis of protein domains to test the contribution of templating mechanisms to epigenetic stability; something definitely could be done with this.*
*I also agree that the idea about the potential role of entanglement in curing the 'small numbers curse' needs further elaboration. The main purpose of this essay was to raise awareness of this very interesting and fundamental physical phenomenon among working biologists, in a hope that some bright young people will come up with new experimental tests for the role of entanglement in living cells. Concerning the 'systems biology' terminology, I had difficulty finding a way to substitute it with another less established word. 'Omics' approaches will most likely be required to address the issue of fluctuations in the context of the cost of maintenance of biological order, and given that 'omics' is a part of systems biology, I consider it appropriate to formulate the discussion in these terms.*

Vlatko Vedral, Quantum Information Group**,** School of Physics and Astronomy, University of Leeds, UK, Professor of Physics at NUS, Singapore.

I am very happy to endorse the paper and recommend publication. Whatever turns out to be the case as far as entanglement in biology is concerned, the exposition is certainly interesting to read. The part on quantum mechanics is very well informed and factually correct, though, of course I cannot judge other parts well. The ideas are definitely fascinating.

*Author's response – I thank Dr. Vedral for his endorsement. As mentioned in my comments to the previous reviewer, one of the goals of this essay is to raise awareness of this interesting phenomenon among biologists, so that they can see alternative ways of how information can be processed in the cells. On the other hand, I think that if entanglement indeed plays a role in life, it might be eventually possible*

*to harness it for the purposes of quantum information processing, but again, it could be done only with the help of biologists.*

Eric Karsenti, Cell Biology and Biophysics unit, EMBL Heidelberg, Heidelberg, Germany.

I don't find anything dramatically wrong in what Ogryzko writes about biology, and I think that the idea is very interesting. I still think that in eukaryotic cells, statistical mechanics and reaction diffusion mechanisms are the most relevant scale to analyze the self- organization of highly dynamic steady state of relatively large structures like the spindle, organelles and overall cell organization. However, it is quite possible that coherence derived from entanglement is also at work underneath the averaging of large numbers of molecules. Obviously, some kind of experimental approaches would be welcome to address this question of entanglement at the cell level.

*Author's response* – *I thank Dr. Karsenti for his comments and agree that experimenters will have the last word in this matter.*

# Endnotes

1. A case in point is the calculation of the number of free protons in a single cell of E.coli [38]. Given the intracellular pH 7.5 and cell volume $\pi/6 \times 10^{-15}$ liters, there are, on the average, only 10 protons in cell [39].

2. A resource, which helps to overcome limitations on what can be achieved by the so called LOCC (Local Operations and Classical Communications) class of processes in quantum information theory [26].

3. The uncertainty due to the coupling with environment can be taken into account using density matrix approach [40].

4. The extra cost can be directly linked to the loss of information about entanglement and estimated using density matrix formalism via comparison of the entropy S(C) of the total system (non-additive quantity in the case of entangled system) with the sum of the entropies of its parts S(A) + S(B).

5. Local Operations and Classical Communications [26].

# Abbreviations

HAT, histone acetyl-transferase; EID, environmentally induced decoherence

# Acknowledgements

The author is grateful to Drs. E. Koonin, V. Vedral and E. Karsenti for the reviews and suggestions how to improve the manuscript. Drs. Murat Saparbaev and Alexander Ishchenko are thanked for many helpful discussions. Author thanks Dr. Linda Pritchard for suggestions how to improve the manuscript and Dr. Marc Lipinski for his support and encouragement.

# Figure legends

**Figure 1. General scheme of epigenetic templating.**

This mechanism implies two essential features: a. Physical linkage between the enzymatic activity depositing a particular epigenetic mark (covalent modification, alternative histone, etc.) and the recognition module for this mark; b. Formation of dimers (or oligomers) by the target of enzymatic activity (protein, nucleic acid, etc.). Shown are three scenarios: I – presence of the mark on one monomer will direct its deposition on the second monomer via recruitment of the depositing activity (can also involve allosteric effects of R-module binding on D-module activity), II – unmarked dimer will not recruit modifier, III – if both monomers are marked, they will not be affected. **R** – recognition module, **D** –deposition module, **T** – target.

**Figure 2. Quantum entanglement.**

**2.A. One example of system setup to observe quantum entanglement**

A pair of entangled particles can be obtained by allowing two previously independent particles to interact and then switching the interaction off. Their spins are measured by observers A and B, who separately choose the angles of the analyzers SG1 and SG2. The entanglement is manifested by the fact that, after several runs of the experiment, both observers obtain random strings of spin values on their respective detectors $D_1$ and $D_2$; however, the correlations between the A and B strings can be seen after direct comparison of the results (**&**). This example illustrates a general and essential feature of an entangled system - *it behaves more predictably than each of its parts*. **SG1**, **SG2** – Stern-Gerlach analyzers, **$D_1$, $D_2$** – detectors, **&** – coincidence monitor.

**2.B. Theoretical explanation**

An entangled state of two particles has to be represented as a linear combination of at least two product states of particle 1 and particle 2. Measurement performed on either one of the particles reduces this superposition to one component, thus redefining the state of the second particle, and influencing the results of its measurement. This is a general property of a composite system with any number of interacting parts, including biological systems.

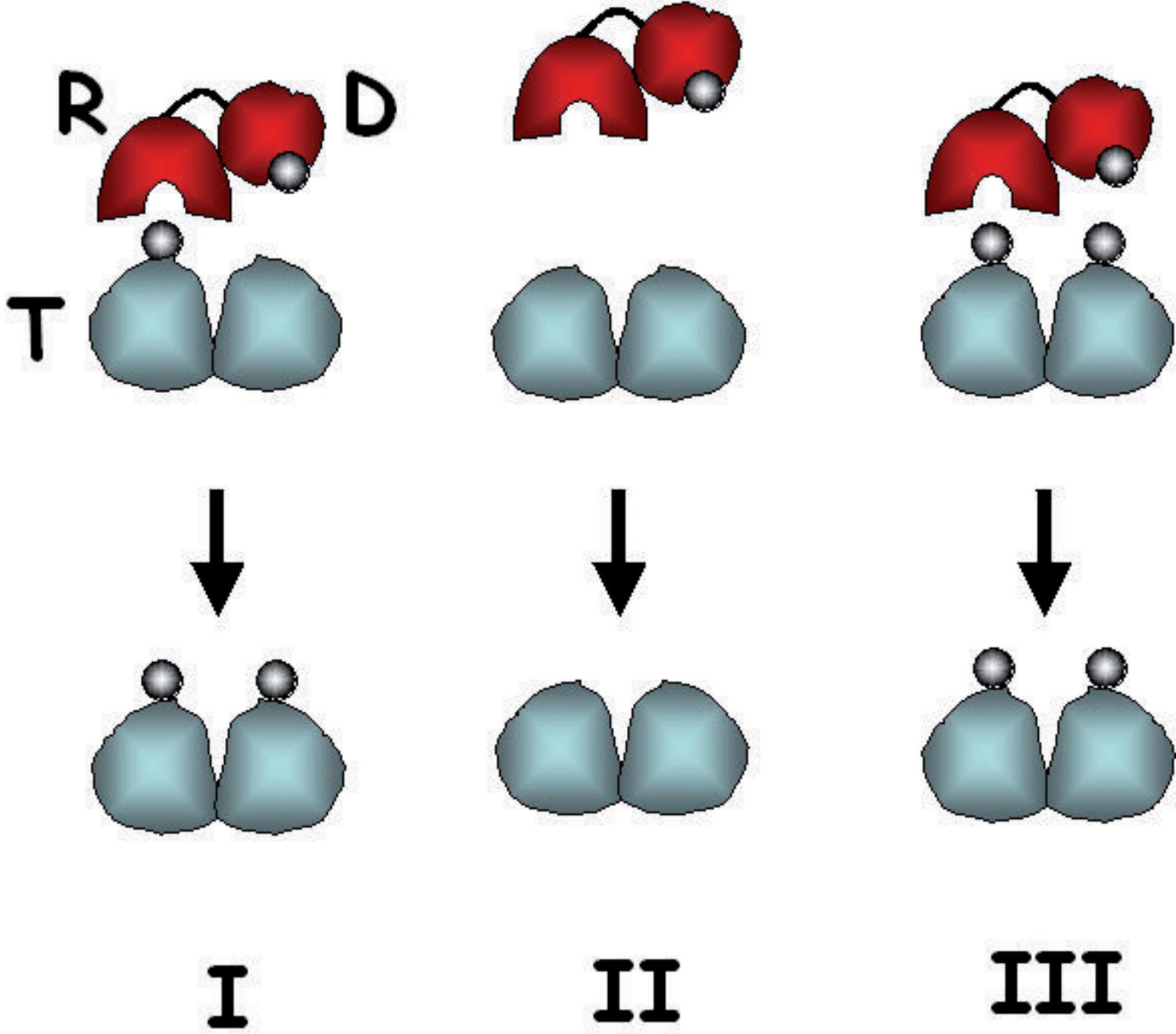

Figure 1

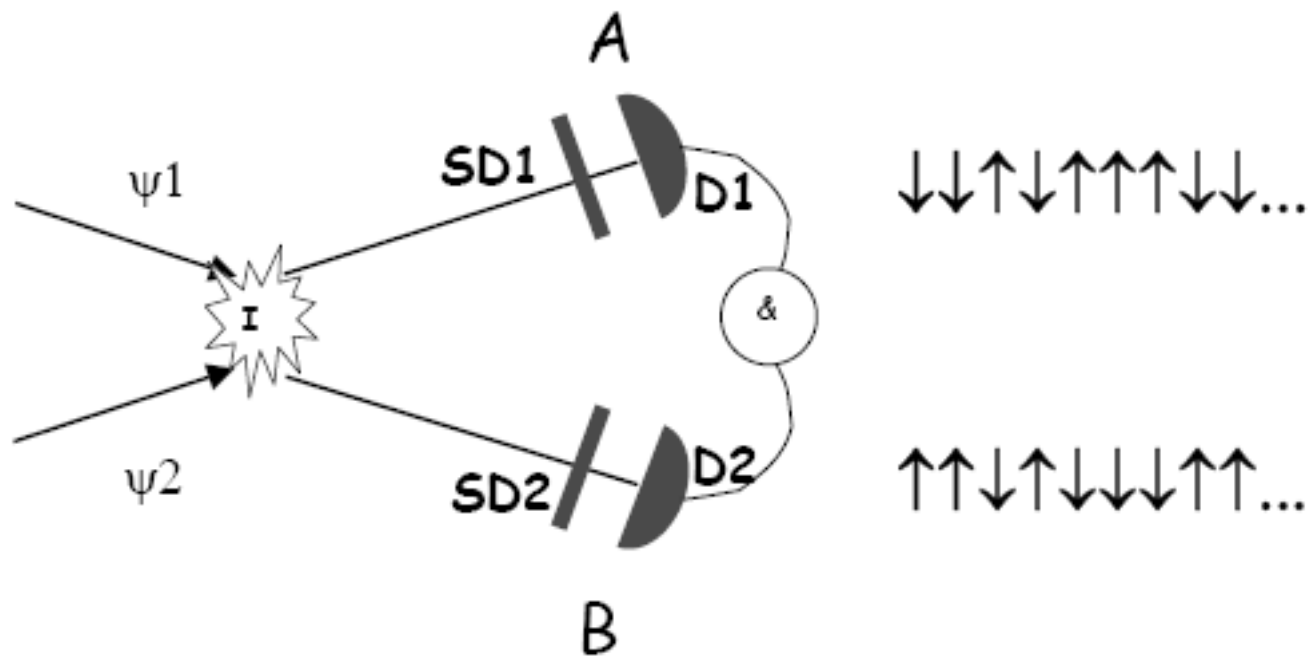

Figure 2